\documentclass[final,5p,times,twocolumn]{elsarticle}
\usepackage{graphicx}
\usepackage{amssymb}
\usepackage{amsmath}
\usepackage{bbm}
\usepackage{bm}
\usepackage{lscape}
\usepackage{color}
\journal{Physics Letters A}
\begin{document}
\begin{frontmatter} 

\title{Recovery of synchronized oscillations on multiplex networks by tuning dynamical time scales}

\author[first]{Aiwin T Vadakkan}
\affiliation[first]{Indian Institute of Science Education and Research Tirupati, Tirupati-517 619, India}
           
\author[second]{Umesh Kumar Verma}
\affiliation[second]{Indian Institute of Technology Indore, Khandwa Road, Simrol, Indore-453 552, India}

\author[third]{G. Ambika\corref{cor1}}
\affiliation[third]{Indian Institute of Science Education and Research Thiruvananthapuram, Thiruvananthapuram-695 551, India}
\cortext[cor1]{Corresponding Author\\ Email address: g.ambika@iisertvm.ac.in}

\begin{abstract} 
 The heterogeneity among interacting dynamical systems or variations in the pattern of their interactions occur naturally in many real complex systems.  Often they lead to partially synchronized states like chimeras or oscillation suppressed states like in-homogeneous or homogeneous steady states. In such cases, it is a challenge to get synchronized oscillations in spite of prevailing heterogeneity.  In this study, we present a formalism for controlling multi layer, multi timescale systems and show how synchronized oscillations can be restored by tuning the dynamical time scales between the layers.  Specifically, we use the model of a multiplex network, where the first layer of coupled oscillators is multiplexed with an environment layer,  that can generate various types of chimera states and suppressed states. We show that by tuning the time scale mismatch between the layers, we can revive the synchronized oscillations. We analyse the nature of the transition of the system to synchronization from various dynamical states and the role of time scale mismatch and strength of inter layer coupling in this scenario. We also consider a three layer multiplex system, where two system layers interact with the common environment layer.  In this case,  we observe anti synchronization and in-homogeneous steady states on the system layers and by tuning their time scale difference with the environment layer, they undergo transition to synchronized oscillations.
\end{abstract}


\end{frontmatter}

\section{Introduction}

The collective phenomena emerging from the interactions of multiple dynamical systems are intensely studied recently and they have a wide range of applications in physics, biology, chemistry, technology, and social sciences. In such studies, the framework of complex networks is effectively utilized to model the complex pattern of interactions \cite{Watts1998}.  However, the variability and heterogeneity of the interacting subsystems and the nature of interactions are best studied using the model of multiplex networks \cite{DeDomenico2013MathematicalNetworks,Kivela2014MultilayerNetworks}. These have wider applications since multiplexing can be between layers of different topology or with different intrinsic dynamics for the subsystems\cite{Verma2020}. Hence, they are found to be most effective in studying biological systems \cite{Boccaletti2009HandbookNetworks, Sadilek2014PhysiologicallyActivity, Makovkin2021}, social interactions \cite{Deville2016ScalingInteractions}, power grids \cite{Pagani2013TheSurvey}, and epidemiology \cite{disease_spread}. 

The emergent dynamics on multiplex networks are rich with several interesting phenomena like diverse forms of synchronization \cite{Synchronization}, cluster synchronization \cite{Pecora2014}, amplitude death \cite{amplitude_death}, and oscillation death \cite{oscillation_death,Explosive_death}. Also, partially synchronized states, like different types of chimera \cite{chimera_pattern,partial_synch,E_Scholl}, form an interesting spatiotemporal behaviour where spatially coherent and incoherent behaviour of states coexist in the system.  Such chimera states emerge due to heterogeneity, either natural or induced, like parameter or frequency mismatch, coupling types \cite{Bogomolov2017},  coupling delay \cite{ZAKHAROVA20157,interplay_delay}, time scale mismatch \cite{Frequency_chimera} and environment coupling \cite{Verma2020}.  Recently, the concept of chimeras is applied to explain many real-world phenomena such as epileptic seizures\cite{Rothkegel2014}, brain activity of various aquatic animals\cite{aquatic_animal} and power grid anomalies \cite{Deng2024}. The wide variety of interesting emergent states related to chimera include amplitude-mediated chimera \cite{Sethia2014ChimeraRevisited}, amplitude chimera \cite{chimera_death_1}, breathing chimera \cite{breathing_chimera}, homogeneous steady state, in-homogeneous steady state, two-cluster steady state, multi-cluster steady state, and chimera death states \cite{chimera_death_1,Banerjee2015} such as one-state chimera death, two-state chimera death, etc.  

The heterogeneity required for inducing chimera states can be explicitly added among interacting systems by considering them on multiplex networks and a few such studies are reported recently \cite{Control_of_Chimera_States}. They include the activation and inhibition of chimera states through multiplexing \cite{PhysRevE.94.052205}, the presence of chimeras in a multi layer structure of neuronal networks \cite{Majhi2017}, and the synchronization of chimeras in multiplex networks \cite{Sawicki2018}. It is reported that weak multiplexing can induce chimera states in neural networks \cite{Semenova2018WeakResonance}. Additionally, the different synchronization scenarios in the multiplex network are widely studied \cite{Synchronization_Induced, Rakshit2019, PhysRevResearch.6.033003}, specifically the effect of inter-layer coupling delay \cite{Sawicki2021, Mao2021DynamicsCouplings}, explosive synchronization \cite{Verma2022,Wu2022DoubleNetworks,collective_phenomena,Bayani2023}, synchronization in adaptive multiplex \cite{Kasatkin2018SynchronizationCouplings} and relay synchronization \cite{Sharma2014EffectOscillators,relay,Rakshit2022}

We note that the two-layer nature of multiplex networks is useful in studying control strategies by which emergent dynamics can be controlled to desired states even when heterogeneity exists. The advantage of control schemes based on multiplexing is that they allow the desired state to be achieved in a certain layer by tuning the other layer which may be more accessible in practice. In this context, controlling the system back to synchronised states from oscillation suppressed states or chimera states will have many applications since synchronised states are desirable for the functionality of many systems like power distribution systems \cite{Motter2013}, brain activity \cite{Fries2001}, coupled laser systems and cardiac cells\cite{Jenkins2013}.

Recent studies report the various collective dynamical states possible in oscillators that interact with each other through a dynamic environment in two-layer multiplex networks. In such systems, due to feedback from the other layer that is nonlocal in connectivity,  a variety of emergent states like amplitude chimera, chimera death states, oscillation suppressed states, multi-cluster steady state etc can occur for different ranges of coupling strengths \cite{Verma2020}. 

Multiple-time-scale phenomena are ubiquitous in nature and observed in temporal neural dynamics \cite{SAMEK2016291,PMID:16116447}, hormonal regulation \cite{PMID:6140270,Radovick1992}, chemical reactions \cite{Das2013}, turbulent flows \cite{Schiestel1987}, and population dynamics \cite{population_dynamics}. When such systems are modeled by complex networks,  the interplay between time scales and the structural properties of the network of nonlinear oscillators can generate many interesting phenomena like amplitude death \cite{Gupta2019}, cluster synchronization \cite{Jalan2016,Yang2017}, and frequency synchronization \cite{Frequency_chimera}.  Among interacting systems, if some systems have slower dynamics than others, their dynamics can be modelled on a multiplex network with the different layers evolving at different time scales.We mention a few such situations; neuronal networks in interaction with glia network, where the glia operates at a slower time scale compared to neurons \cite{Gong2024}, environmental effects on ecosystems, where the changes in environment can be slower \cite{Feudel2023} and power transmission networks with variations in the subsystems and their frequencies \cite{Strenge2020}. Most of the studies on synchronization in multiplex networks are on layers with identical dynamical time scales. The effect of difference in time scales between the layers in deciding their emergent states is still unexplored.

 In this study we consider the evolution of emergent states in a multiplex network with heterogeneity in dynamics and interactions and present how synchronized oscillations  can be restored by tuning the dynamical time scales of the system. We illustrate this using the model of a multiplex network where the system layer of coupled oscillators is multiplexed with an environment layer.  In studies reported so far,  the time scale mismatch among systems on a network  is shown to drive the systems to suppression of oscillations\cite{Gupta2019}.  But we show that by tuning the time scale mismatch between the layers, we can restore synchronized oscillations in both layers from chimera or death states.  We analyse the nature of the transition to synchronization, and the results are verified for two- and three-layer multiplex networks, with the  Stuart-Landau (SL) oscillator as the system dynamics controlled by the environment layer. 

\section{Two layer system}

We start by considering a two layer multiplex system, with one layer comprising of a ring network of SL oscillators, referred to as the system layer, L1, which has local intra layer diffusive coupling. The second layer L2, which is the environment layer, is  modeled as a ring network of 1-d over damped oscillators with nonlocal intra layer diffusive couplings. The $N$ oscillators in L1 are connected to tho corresponding ones in L2 via interlayer coupling of the feedback type to form a multiplex network. The dynamics of the environment L2 is sustained by the negative feedback from L1, and the  dynamics of L1 is controlled by L2 via its positive feedback coupling. The equations for the dynamics of such a two-layer multiplex network are as given below.

\begin{eqnarray}
\dot{x}_{i}&=&(1-x_i^2-y_i^2)x_i-\omega y_i+\frac{K_1}{2P_1}\sum_{j=i-P_1}^{i+P_1}(x_j-x_i)+\epsilon s_i\nonumber\\
\dot{y}_{i}&=&(1-x_i^2-y_i^2)y_i+\omega x_i \nonumber\\
\dot{s}_{i}&=&\tau[-\gamma s_i -\epsilon x_i +\frac{K_2}{2 P_2}\sum_{j = i-P_2}^{i+P_2} (s_j-s_i)]
\label{eq1}  		
\end{eqnarray}

where the dynamics of SL oscillators in layer L1 are characterized by the variables $x_i$ and $y_i$ for i = 1, 2, ..., N, and  $\omega$ is the natural frequency of their intrinsic limit cycle oscillations. The 1-d over damped oscillators, $s_i$, in L2 have a positive damping coefficient, $\gamma$. The interaction between the layers is through feedback coupling of strength $\epsilon$.  The interactions among the SL oscillators in the first layer are regulated by  the intra-layer coupling strength, $K_1$, and coupling range, $P_1$, while that of the environment is controlled by $K_2$ and $P_2$. Here $P$ corresponds to the number of nearest neighbors in each direction; hence, $P \in \{1, \frac{N}{2}\}$, with $P=1$ for local connections, $P=\frac{N}{2}$ for a global coupling, for other cases $P$ is in the range $1 < P < \frac{N}{2}$. The parameter $\tau$ is introduced to represent the mismatch in dynamical time scales between the two layers. Thus a value of $\tau < 1$ means the environment layer L2  evolves at a slower time scale compared to layer L1.

Initially we keep $\tau = 1$ so that both layers evolve with the same time scale. Following the recent study \cite{Verma2020},  we generate various emergent dynamical states in the system by varying the value of $\epsilon$, such as  amplitude chimera(AC), homogeneous steady state(HSS), in homogeneous steady state(IHSS), one-state chimera death(1-CD), two-state chimera death(2-CD) etc. For this, we choose appropriate cluster-like initial conditions, for the $N=100$ oscillators in L1,  while the systems in L2 start with random initial conditions\cite{Verma2020}.

\begin{figure*}
 \includegraphics[width=0.9\textwidth]{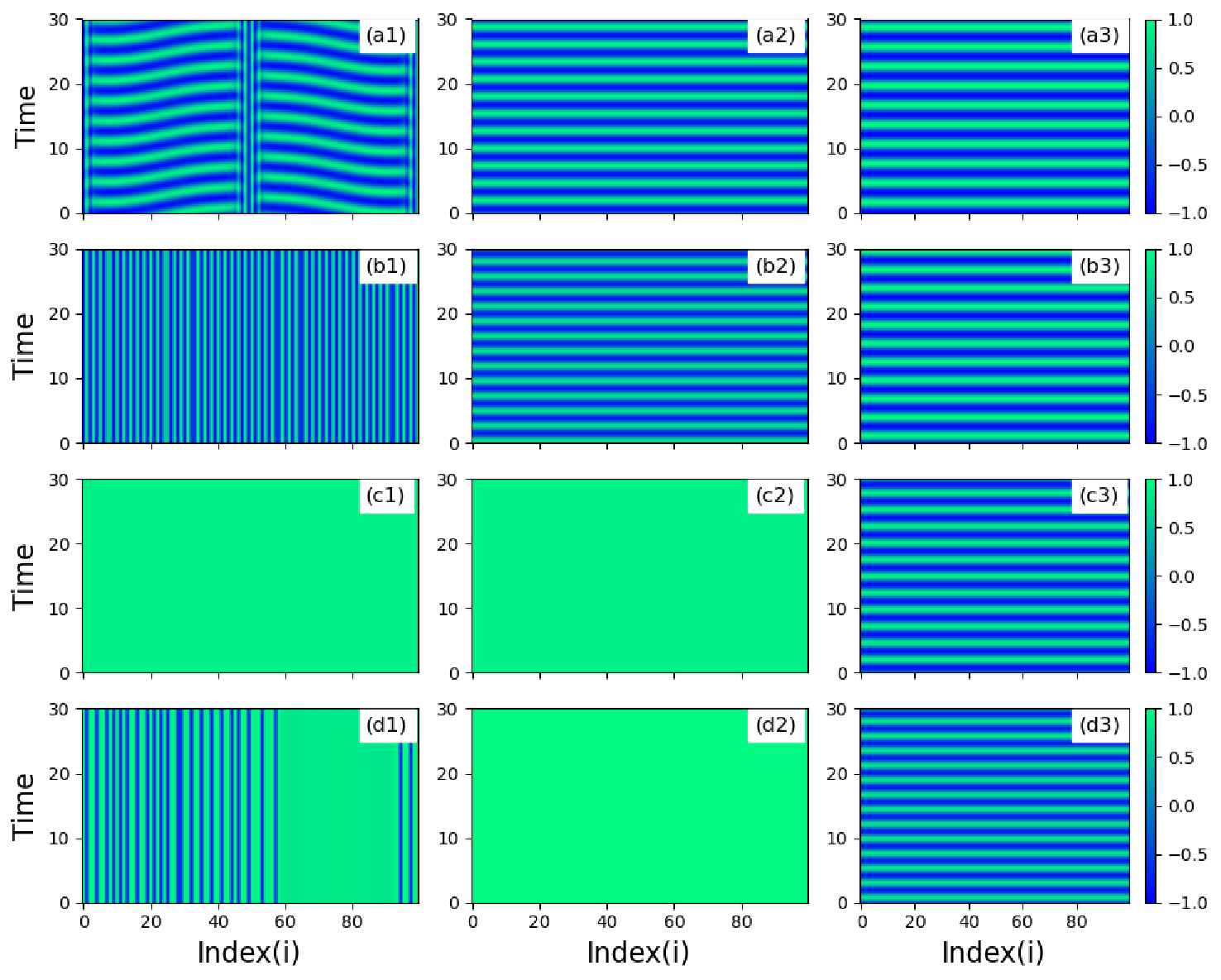}
 \caption{Spatio temporal dynamics for different values of $\epsilon$ and $\tau$. (a1): Amplitude chimera at $\epsilon=2$ and $\tau=1$, (a2) and (a3): Synchronized oscillations at $\epsilon=2$, $\tau=0.4$ and $0.1$ respectively, (b1): Inhomogeneous steady state at $\epsilon=3$ and $\tau=1$, (b2) and (b3): Synchronized oscillations at $\epsilon=3$ and $\tau=0.4$ and $0.1$, (c1) and (c2): Homogeneous steady state at $\epsilon=4.5$, $\tau=1$ and $0.4$, (c3): Synchronized oscillations at $\epsilon=4.5$ and $\tau=0.1$, (d1): One state chimera death at $\epsilon=6.15$ and $\tau=1$, (d2): Homogeneous steady state at $\epsilon=6.15$ and $\tau=0.4$ and (d3): Synchronized oscillations at $\epsilon=6.15$ and $\tau=0.1$. The other parameter values used are : $K_1=K_2=10$, $P_1=1$, $P_2=25$, $\omega=2$ and $\gamma=1$.By adjusting the time scale difference between the layers, synchronised oscillations are seen to be restored from the various emergent states considered.}
 \label{fig1}
 \end{figure*}
 
\subsection{Recovery of synchronized oscillations}

In this section, we present the role of time scale separation between the layers in controlling the emergent dynamics of the system. Specifically, we show how introducing a disparity in the time scale between the layers can facilitate the recovery of synchronized oscillations from the above-mentioned dynamical states. We consider the scenario in which all SL oscillators in L1 are connected with one another locally, $P_1 = 1$. At the same time, the environment is coupled nonlocally, with $P_2 = 25$. We set the values of $K_1$ and $K_2$ to 10, and investigate the impact of varying $\tau$ for different values of interlayer coupling strengths, $\epsilon$.

The spatiotemporal plots from the $y$ variable of SL oscillators in L1 for different values of $\epsilon$ and $\tau$ are shown in Fig.~\ref{fig1}. As shown in Fig.~\ref{fig1}(a1) when $\epsilon=2$ and $\tau=1$, the system exhibits a stable amplitude chimera, which is characterized by the coexistence of coherent and incoherent oscillations with respect to amplitude. The system persists in this state as $\tau$ is reduced from 1 up to 0.5. But as the mismatch increases or the $\tau$ value is further reduced, synchronized oscillations are restored in both the layers. This is clear from Fig.~\ref{fig1}(a2) and Fig.~\ref{fig1}(a3), plotted for the same $\epsilon$ value but with $\tau=0.4$ and 0.1 respectively. While both show synchronized oscillations, we find the frequency of oscillations in L1 depends on $\tau$.  When $\epsilon=3$ and $\tau=1$, the system emerges into  an IHSS, (Fig.~\ref{fig1}(b1)), from which synchronized oscillations are revived by reducing $\tau$ as shown in Fig.~\ref{fig1}(b2,b3). The IHSS is seen for $\tau$ values ranging from 1 to 0.6, after which synchronized oscillations are revived. For $\epsilon=4.5$ and $\tau$ values ranging from 1 to 0.4, HSS is observed in the system as shown in Fig.~\ref{fig1}(c1,c2) and for smaller values of $\tau$ synchronized oscillations are revived(Fig.~\ref{fig1}(c3)). Similarly for $\epsilon=6.15$ and $\tau$ in the range 1.0 to 0.5, the emergent state is one state chimera death(1-CD), (Fig.~\ref{fig1}(d1)). For $\tau$ values ranging from 0.4 to 0.2, the system settles to HSS as in Fig.~\ref{fig1}(d2). For $\tau$=0.1 synchronized oscillations are restored in the system as shown in Fig.~\ref{fig1}(d3). Thus the mismatch in time scales required to revive synchronized oscillations depends on the strength of interlayer connections parameterised by $\epsilon$. 

To check the stability of the revived oscillations, we add white noise of strength 0.5 for a short time of 10 time steps  as an external perturbation. We observe the system returns to the synchronized oscillations shortly after the noisy perturbation is removed. We have also verified the results with random initial conditions on both layers. For instance,  for values of $\epsilon=5$ and $\tau=1$, IHSS is observed in the system. As $\tau$ is reduced the systems goes to HSS which is followed by the revival of synchronized oscillations. 

Furthermore, the study is extended with different nodal dynamics on system layer L1 , such as the van der Pol oscillator with limit cycle oscillations and the Rössler system with chaotic dynamics. In the case of the van der Pol oscillator, for random initial conditions, with the same setting as the previous case, IHSS is observed for $\epsilon=5$, from which synchronized oscillations are recovered reducing $\tau$.

\begin{figure}[!ht]
 \includegraphics[width=0.48\textwidth]{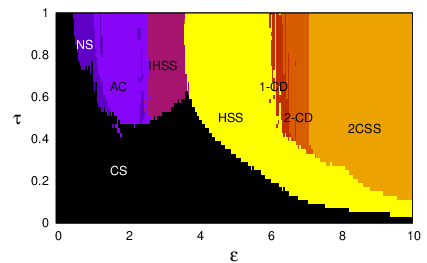}
 \caption{Parameter plane, $\epsilon$ vs $\tau$, for the two-layer multiplex network with Stuart-Landau oscillators in L1 and the environment in L2 for $K_1=K_2=10$, $P_1=1$, $P_2=25$, $\omega=2$ and $\gamma=1$. Here, the dynamical states seen are, CS - Complete synchronization, NS - No-synchronization, AC - Amplitude chimera, IHSS - Inhomogeneous steady state, HSS - Homogeneous steady state
1-CD - One-state chimera death, 2-CD - Two-state chimera death, and 2CSS - Two-cluster steady state. We see the combined effect of the time scale mismatch and the intralayer coupling strength on the transition to synchrony.}
 \label{fig2}
 \end{figure}

We also study emergent dynamics in a two-layer multiplex network with $N=100$ chaotic Rössler oscillators in L1 coupled with the environment in L2, keeping $P_1=P_2=10$, $K_1=K_2=5$, and $\gamma=1$.
We observe both L1 and L2 in synchronized chaotic state for small values of $\epsilon$. As $\epsilon$ is increased, the system undergoes reverse period-doubling transitions to reach HSS. However, by introducing a time scale mismatch, the system can revert back to synchronized chaotic behaviour from HSS or from any periodic cycle.

\begin{figure}[!ht]
 \includegraphics[width=0.48\textwidth]{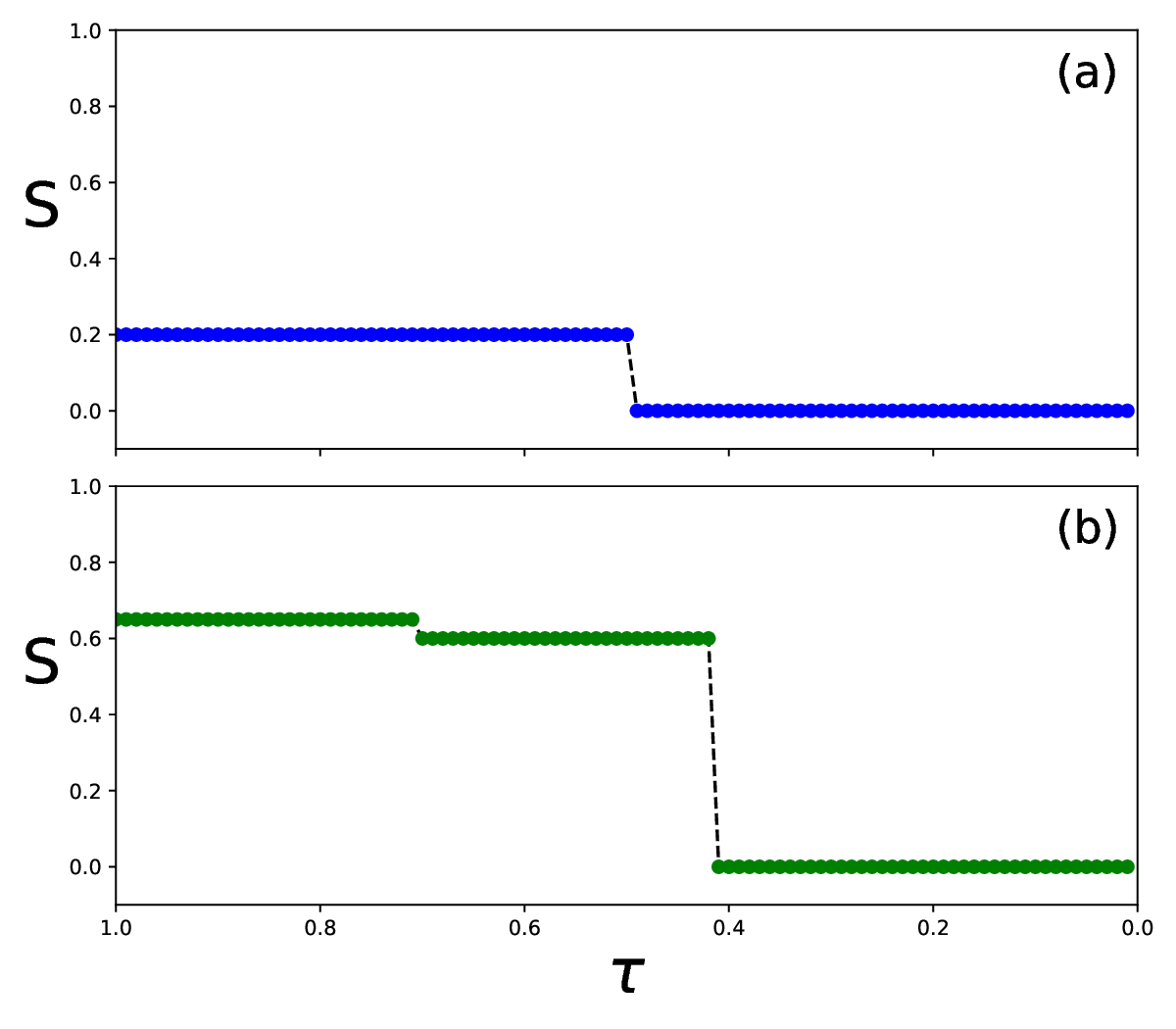}
 \caption{Strength of incoherence ($S$) with decreasing $\tau$ indicating the transition to synchronized state from (a): amplitude chimera at $\epsilon=2$, and (b): one state chimera death at $\epsilon=6.15$. The other parameter values are:  $K_1=K_2=10$, $P_1=1$, $P_2=25$, $\omega=2$ and $\gamma=1$. In each case, the transition from the chimera state to synchronization occurs at a critical value of $\tau$.}
 \label{fig3}
 \end{figure}

\subsection{Characterization of the dynamical states and their
transitions}

To characterise the nature of the emergent dynamics and study the transitions in the dynamics of the system, we calculate the strength of incoherence (S), as introduced by Gopal et al\cite{strength_incoherence}.  By definition, $S=0$ for the spatially synchronized state, $S=1$ for the desynchronized state and an intermediate value between 0 and 1 indicates the chimera or the cluster states. 

We begin by determining $w = x_{i}-x_{i+1}$, where $i$ represents the index of the oscillators. By grouping the oscillators into M bins of equal size $n$, such that $n=\frac{N}{M}$, the local standard deviation, $\sigma(m)$, is defined as

\begin{equation}
    \sigma(m) = \Biggl \langle \sqrt{\frac{1}{n}\sum_{j=n(m-1)+1}^{mn}[w_{j} - \bar{w}_{j}]^2}\Biggl \rangle _t
\end{equation}

where $\bar{w} = \frac{1}{n}\sum_{j=n(m-1)+1}^{mn} w_{j}(t)$  and $\langle..\rangle_{t}$ represents average over time. Now,  $S$ is obtained as, 

\begin{equation}
    S = 1 - \frac{\sum_{m=1}^M s_m}{M}, \quad  s_m= \Phi(\delta-\sigma(m))
\end{equation}

The Heaviside step function $\Phi(.)$ of $\sigma(m)$ and $\delta$ is used to calculate $s_m$, with $\delta$ being a predefined threshold value set as a percentage of the difference between the maximum and minimum values of $x$. In our calculations, we set $\delta=0.2$ and $M=20$. 

To track the transition from a state of suppressed oscillations, we compute the  average amplitude as an order parameter as shown below.

\begin{equation}
    <A(\tau)>=\frac{1}{N}\sum_{i=1}^{N}[\langle x_{i,max} \rangle_t-\langle x_{i,min} \rangle_t]
\end{equation}
The value of $A(\tau) = 0$ when all oscillators are in the stationary state of HSS or IHSS, whereas when they are in the oscillatory state, the value of $A(\tau) > 0$.

Using the above two measures, we map the spatiotemporal behavior of the two layer multiplex system for a range of the parameters,  $\epsilon$ and $\tau$ with the other parameter values kept as $K_1=K_2=10$, $P_1=1$, $P_2=25$, $\omega=2$ and $\gamma=1$.  We mark the variety of emergent states possible on the parameter plane of $\epsilon$ vs $\tau$ in Fig.~\ref{fig2}.  The regions of distinct dynamical states shown include complete synchronization(CS), no synchronization(NS), chimera states such as amplitude chimera(AC), one state chimera death(1-CD), two state chimera death(2-CD), various steady states such as homogeneous steady state(HSS), inhomogeneous steady state(IHSS) and two cluster steady state(2-CSS). It is clear from Fig.~\ref{fig2} that as $\tau$ decreases, the region of synchronized oscillations expands, indicating the system's transition from various dynamical states to synchronized oscillations. The phase  diagram in Fig.~\ref{fig2} also reveals the combined effect of the time scale mismatch $\tau$ and the intralayer coupling strength $\epsilon$ on the transition to intralyer synchrony.

We use the strength of incoherence $S$ to identify nature of the transition from chimera state to complete synchronization. We illustrate this for two such transitions in Fig.~\ref{fig3}, where $S$ is plotted against $\tau$  for $\epsilon=2$(a) and $\epsilon=6.15$(b).  The variation in $S$ indicates the tipping of the system to the state of synchronized oscillations from  amplitude chimera and one state chimera death respectively.

\begin{figure}[!ht]
    \includegraphics[width=0.48\textwidth]{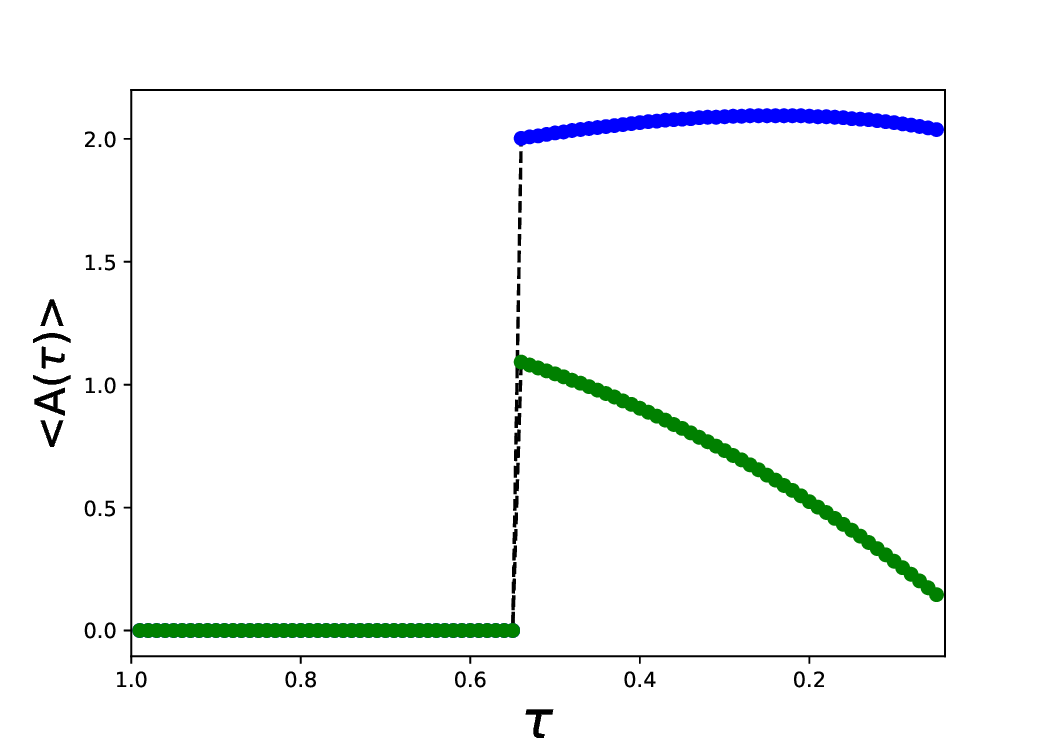}
    \caption{Average amplitude with decreasing $\tau$ for $\epsilon=3$. The sudden transition from inhomogeneous steady state(IHSS) to synchronized oscillations is seen in layer L1(blue) and layer L2(green), as the $\tau$ is varied. The other parameter values are:  $K_1=K_2=10$, $P_1=1$, $P_2=25$, $\omega=2$ and $\gamma=1$.}
    \label{fig4}
\end{figure}

Further, we study the variation of the average amplitude of the system  with decreasing $\tau$ or increasing mismatch in time scale as plotted in Fig.~\ref{fig4} for the specific case of $\epsilon=3$. We compute $<A(\tau)>$ starting from cluster like initial conditions for each value of $\tau$. By decreasing $\tau$ from 1.0, we look for the transition to synchronized state from the state of IHSS. For $\tau=1$, the system is in IHSS with the average amplitude of zero. As $\tau$ is reduced, we see an abrupt increase in value of $A$ indicating the sudden transition to large amplitude synchronous oscillations for both layers for a critical value of $\tau$.

\begin{figure}
\centering
\includegraphics[width=0.48\linewidth]{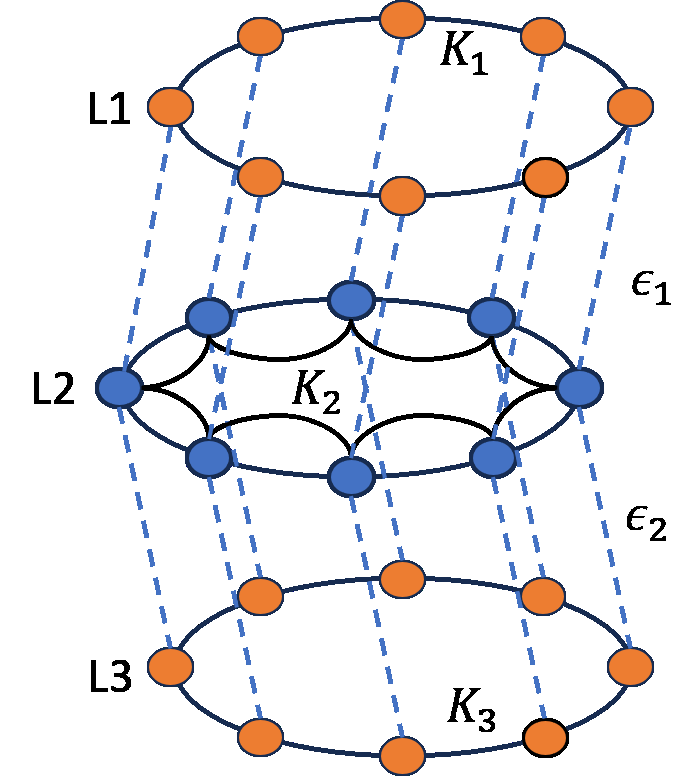}
 \caption{Schematic representation of the three-layer network with  $N=8$, where nodes in the middle layer (L2) (blue) model the environment while nodes in the upper (L1) and lower (L3) layers (orange) represent the system layers.  L1 and L3 have local and L2 nonlocal intralayer connections of strength $K_i$ , while interlayer connections are feedback type of strength $\epsilon_i$.}
\label{fig5}
\end{figure}

\section{Three layer system}

The study is extended to a three layer multiplex system with the L1 and L3 layers as networks of $N$ Stuart-Landau oscillators and the environment in the middle layer L2 as shown in the Fig.~\ref{fig5}. The equations for the dynamics of the three layer system are presented below.

\small \begin{eqnarray}
\dot{x}_{i1}&=&(1-x_{i1}^2-y_{i1}^2)x_{i1}-\omega y_{i1}+\frac{K_1}{2P_1}\sum_{j=i-P_1}^{i+P_1}(x_{j1}-x_{i1})+\epsilon_1 s_{i2}\nonumber\\
\dot{y}_{i1}&=&(1-x_{i1}^2-y_{i1}^2)y_{i1}+\omega x_{i1} \nonumber\\
\dot{s}_{i2}&=& \tau[-\gamma s_{i2} -\epsilon_1 x_{i1}-\epsilon_2 x_{i3} +\frac{K_2}{2 P_2}\sum_{j = i-P_2}^{i+P_2} (s_{j2}-s_{i2})] \nonumber\\
\dot{x}_{i3}&=&(1-x_{i3}^2-y_{i3}^2)x_{i3}-\omega y_{i3}+\frac{K_3}{2P_3}\sum_{j=i-P_3}^{i+P_3}(x_{j3}-x_{i3})+\epsilon_2 s_{i2} \nonumber\\
\dot{y}_{i3}&=&(1-x_{i3}^2-y_{i3}^2)y_{i3}+\omega x_{i3} \nonumber\\
\label{eq2}  		
\end{eqnarray}

\begin{figure}

\includegraphics[width=0.48\textwidth]{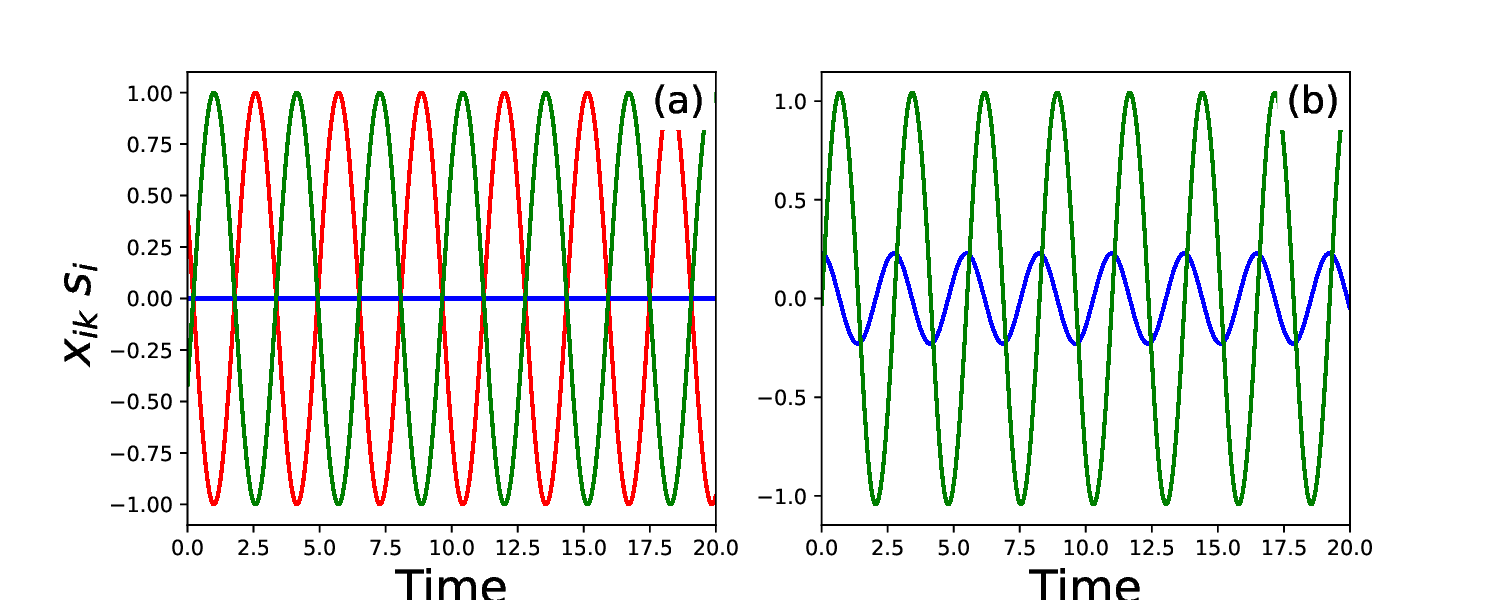}
\caption{Time series of the x-variable of Stuart-Landau oscillators (L1-red, L3-green) and s-variable of the environment (blue). (a): Anti synchronization between L1 and L3 for $\epsilon_1=\epsilon_2=2.5$ and $\tau=1$, (b): Complete synchronization between L1 and L3 for $\epsilon_1=\epsilon_2=2.5$ and $\tau=0.1$. The other parameter values are: $K_1=K_2=K_3=15$, $P_1=P_3=1$, $P_2=25$, $\omega=2$ and $\gamma=1$. It is seen that as $\tau$ is decreased, the two layers undergo transition from anti synchronization to synchronized oscillations.}
\label{fig6}
\end{figure}

\begin{figure}
    \includegraphics[width=0.48\textwidth]{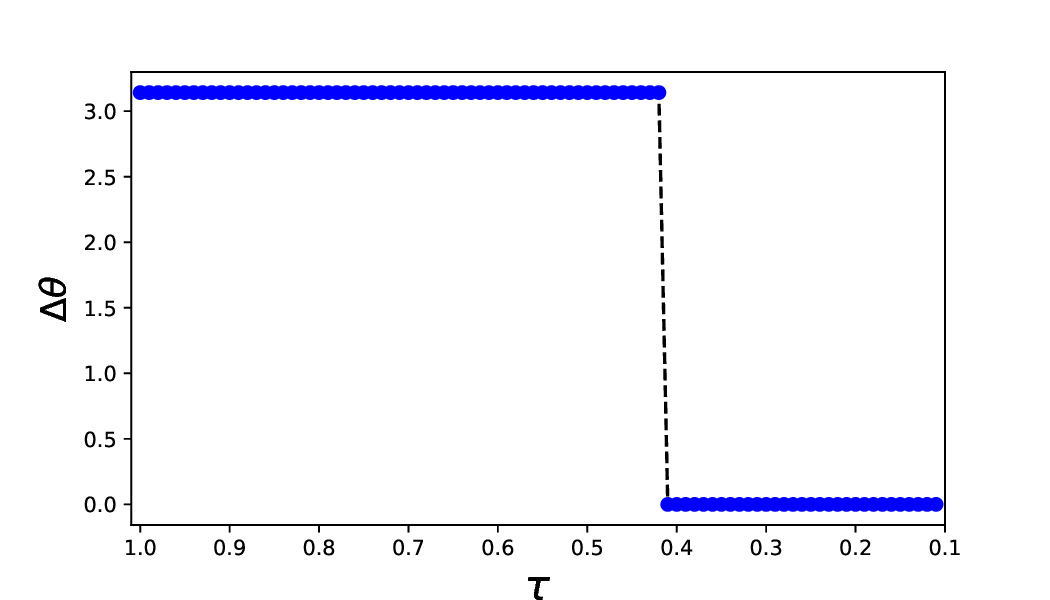}
    \caption{Variation in the phase difference $\Delta \theta$ between L1 and L3 as $\tau$ is varied for $\epsilon_1=\epsilon_2=2.5$.  The other parameter values are: $K_1=K_2=K_3=15$, $P_1=P_3=1$, $P_2=25$, $\omega=2$ and $\gamma=1$. As $\tau$ is reduced, we see synchronization is revived from IHSS.}
    \label{fig7}
\end{figure}

where the SL oscillators are characterized by the variables $x_{ik}$ and $y_{ik}$ for i = 1, 2, ..., N in the $k^{th}$ layer, where $k=1,3$. In addition, $\epsilon_1$ and $\epsilon_2$ represent the inter-layer coupling strength for L1 and L3, respectively. All other parameters are consistent with those previously described in the two-layer scenario.

\begin{figure}

    \includegraphics[width=0.48\textwidth]{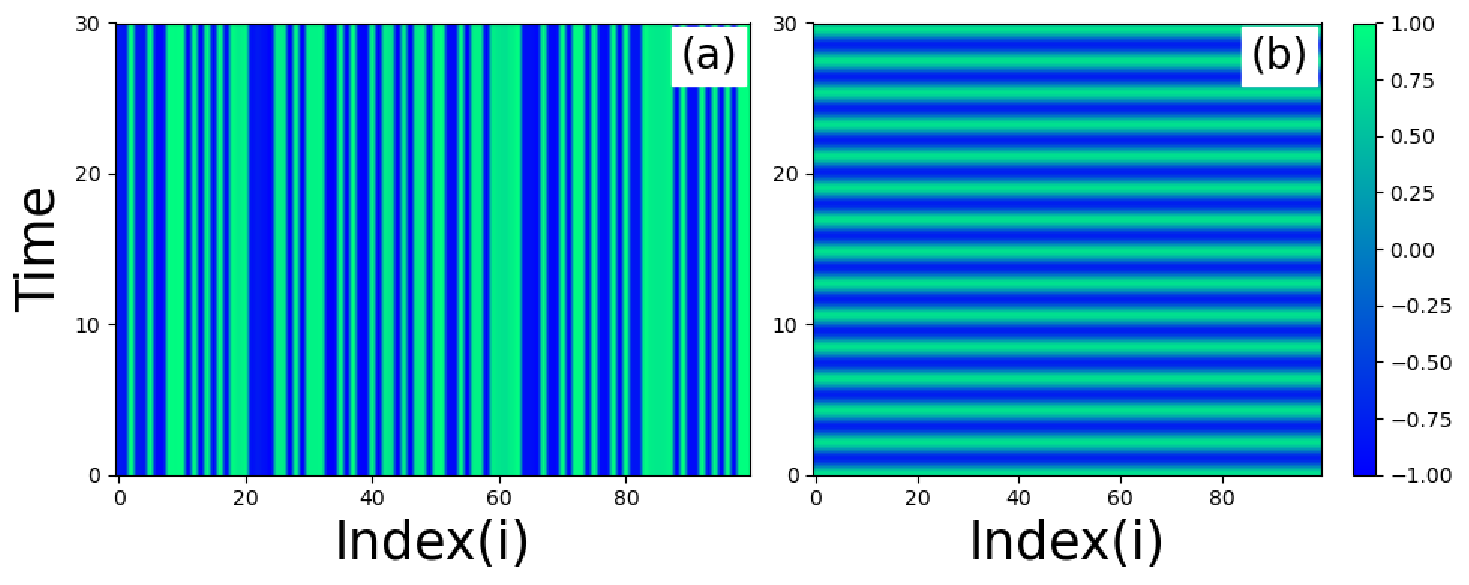}
    \caption{Spatio temporal plots from the $y_i$ variable of Stuart-Landau oscillators in L1  for the three-layer multiplex network. (a): Inhomogeneous steady state for $\epsilon_1=\epsilon_2=5$ and $\tau=1$, (b): Synchronized oscillations for $\epsilon_1=\epsilon_2=5$ and $\tau=0.1$.  The other parameter values are: $P_1=P_3=1$, $P_2=40$, $K_1=K_3=30$, $K_2=1$, $\omega=2$ and $\gamma=1$. As $\tau$ is reduced, we see synchronization is revived from IHSS.}
    \label{fig8}
\end{figure}

Similar to the two-layer case, we consider all SL oscillators in L1 and L3 connected with one another locally, with $P_1= P_3 = 1$ and the environment L2 is coupled nonlocally, with $P_2=25$. When there is no time scale mismatch between the layers, for random initial conditions we observe synchronized oscillations in the system layers, but with L1 and L3 being anti synchronized with each other for parameter values of $K_1=K_2=K_3=15$, $\gamma=1$, $\omega=2$ and $\epsilon_1=\epsilon_2=2.5$ as shown in Fig.~\ref{fig6}(a). In this case, the environment goes to death due to the antiphase feedback from the system layers. As time scale mismatch is introduced, the layers L1 and L2 transit from anti synchronization to complete synchronization, as shown in Fig.~\ref{fig6}(b) with the revival of oscillations in L2 also. In Fig.~\ref{fig7} we show this transition by plotting the phase difference $\Delta \theta$ between L1 and L3, as $\tau$ is decreased. 

A similar recovery of synchronized oscillations is observed from a state of IHSS also. For parameter values of $P_1=P_3=1$, $P_2=40$, $K_1=K_3=30$, $K_2=1$, $\gamma=1$ and $\omega=2$, the system shows IHSS for $\epsilon_1=\epsilon_2=5$ and $\tau=1$ as shown in Fig.~\ref{fig8}(a) from which synchronized oscillations are revived as depicted in Fig.~\ref{fig8}(b) for $\tau=0.1$. For the three layer system, with $\tau=1$, chimera states are seen only for very specific initial conditions and they undergo transitions to synchronization as $\tau$ is decreased. Thus we show that it is possible to revive synchronized oscillations from different dynamical states by tuning the parameter $\tau$ in the three layer system also. 

\section{Conclusion}

While there are several studies on the collective dynamics of multiplex networks, the role of time scales remains unexplored. In this letter, we discuss how the time scale differences between layers can be exploited to  revive the synchronized oscillations in a multiplex network of nonidentical layers. When we consider each layer evolving at a different time scale, we can introduce the time scale mismatch between the layers as an effective parameter that can control the dynamics of the system. 

We first consider a two-layer network where the first layer L1 consists of an ensemble of Stuart-Landau (SL) oscillators, while the second layer L2 model the environment. When both layers evolve with the same time scale, the coupled system exhibits various emergent phenomena, such as amplitude chimera, chimera death, HSS, IHSS, and 2-CSS. By tuning the time scale difference between the layers, we show the synchronized oscillations can be restored in both layers from these different states. By computing the measures, strength of incoherence and average amplitude, we characterise the possible dynamical states and show the nature of their transitions to synchronized oscillations. The revival of synchronized oscillations are also shown with van der Pol oscillators and chaotic Rössler systems on layer L1. 

We also consider a three layer multiplex network, with the upper and lower layers, L1 and L3,  as coupled SL oscillators, and the middle layer, L2, as the environment. The most dominant emergent states in this case are anti-synchronization between L1 and L3 and IHSS in both, for random and cluster like initial conditions, depending on $\epsilon$ and extent of nonlocal connections in L2.  For identical time scales and lower value of interlayer coupling, first layer's oscillators are exactly in anti-phase with third layer oscillators. When mismatch is time scale is introduced between the environment and system layers, the system layers go to complete synchronization. Similarly for identical time scale but higher value of interlayer coupling, the system shows inhomogeneous steady state from which synchronized oscillations are revived by introducing time scale difference between layers.  
Thus in two layer and three layer multiplex systems studied, the main result is the revival of synchronization, but the synchronization in the three layer case is achieved remotely between L1 and L3 by tuning the time scale of the shared common layer L2. 

We note real-world  complex systems exist where different time scales and structural patterns coexist resulting in a variety of emergent states like in neuronal networks, power grids, climate and ecosystems. The present study extends our understanding of emergent states of such systems modeled using the framework of multiplex networks, with different layers evolving at nonidentical dynamical time scales. The model introduced here is highly flexible and has relevance in controlling systems evolving under differing time scales to desired synchronized oscillations. Thus in eco systems, the time scale difference between intrinsic dynamics and environmental changes is shown to have a decisive role in inducing transitions in the system \cite{Feudel2023}. Here the role of a slowly changing environment is studied by introducing a drift in the relevant parameters. In our study we include environment as a layer multiplexed with the system layer, which  gives more flexibility to track the impact of slowly varying environment. This also provides a way to arrive at possible control mechanisms to achieve desired emergent states even in the presence of heterogeneity among interacting systems.

\end{document}